\documentclass[aps,pra,showpacs,amsmath,amssymb,twocolumn]{revtex4-1}

\usepackage{graphicx}
\usepackage{dcolumn}
\usepackage{bm}
\usepackage{enumerate}
\usepackage{amsmath}
\usepackage{amssymb}
\usepackage{hyperref}
\usepackage{color}
\usepackage{times}

\allowbreak
 
\begin{document}

\title{Finite size effects on topological interface states in one-dimensional scattering systems
}

\author{P.~A.~Kalozoumis}
\email[]{pkalozoum@phys.uoa.gr}
\affiliation{ Laboratoire d'Acoustique de l'Universit\'e du Maine, UMR CNRS 6613 Av. O. Messiaen, F-72085 LE MANS Cedex 9, France}
\affiliation{Department of Physics, University of Athens, 15771 Athens, Greece}

\author{G.~Theocharis}
\affiliation{ Laboratoire d'Acoustique de l'Universit\'e du Maine, UMR CNRS 6613 Av. O. Messiaen, F-72085 LE MANS Cedex 9, France}

\author{V.~Achilleos}
\affiliation{ Laboratoire d'Acoustique de l'Universit\'e du Maine, UMR CNRS 6613 Av. O. Messiaen, F-72085 LE MANS Cedex 9, France}

\author{S.~F\'elix}
\affiliation{ Laboratoire d'Acoustique de l'Universit\'e du Maine, UMR CNRS 6613 Av. O. Messiaen, F-72085 LE MANS Cedex 9, France}

\author{O.~Richoux}
\affiliation{ Laboratoire d'Acoustique de l'Universit\'e du Maine, UMR CNRS 6613 Av. O. Messiaen, F-72085 LE MANS Cedex 9, France}

\author{V.~Pagneux}
\affiliation{ Laboratoire d'Acoustique de l'Universit\'e du Maine, UMR CNRS 6613 Av. O. Messiaen, F-72085 LE MANS Cedex 9, France}

\date{\today}

\begin{abstract}
One-dimensional topological edge modes are usually studied considering the interface between two different semi infinite periodic crystals (PCs) with inverted band structure around the Dirac point. Here we consider the case where the two PCs are finite, constituting an open scattering system, and we study the influence of the size of this finite structure on the interface mode by inspecting the complex resonances. First we show the complex resonance distribution corresponding to the band inversion around the Dirac point. Perturbations from the Dirac point display the emergence of the localized interface mode. We also report on a remarkable robustness of the interface mode as the system size varies which persists even for the smallest possible size.
\end{abstract}


\maketitle

\section{Introduction}\label{sec1}

In the past decade topological insulators~\cite{HasanRevModPhys2010,KanePRL2005} have developed into a field of intense research activity, attracting attention from diverse fields in Physics ranging from superconductors~\cite{QiRevModPhys2011,AnindyaNatPhys2012} to acoustics~\cite{XiaoNatPhys2015}. 
The presence of energy gaps in the band structure of their bulk Hamiltonian renders themTopologically Protected Qubits from a Possible Non-Abelian Fractional Quantum Hall State
insulators for these energy ranges. However, when suitable interfaces are formed, they can exhibit conducting surface states with energies residing within certain energy gaps. 
A key feature of interface states which emerge in topological insulators is their \textit{topological protection} and consequently their robustness in the presense of defects and perturbations, as long as the band gap is preserved. Topologically protected interface states have been reported in a multitude of systems including graphene~\cite{GailPRB2011,ZhangPRL2011}, condensed matter models with chiral edge states~\cite{BurkovPRL2011, RoushanNat2009}, quantum Hall systems~\cite{SarmaPRL2005,KonigScience2007} and the Su-Schrieffer-Heeger (SSH) model~\cite{KaneNatPhys2014,SchomerusOptLett2013} for polyacetylene. The properties of such protected surface states have been associated with the existence of topological invariants related to the bulk band structure such as the
Chern number~\cite{ThoulessPRL1982,MoorePRB2007} and the Zak phase~\cite{ZakPRL1989} for 1D systems. 

There is a growing interest in  the photonic counterparts of topological insulators~\cite{JoannopoulosNatPhot2014,KhanikaevNatMat2013} as they also possess analogous protected surface wave states. In the case of discrete optics, topological properties have been addressed via discrete waveguide systems described by the coupled mode theory~\cite{RechtsmanNat2013,WeimannNatMat2017,SchomerusOptLett2013}. 
Recently, continuous one-dimensional (1D) structures, with topologically protected interface states~\cite{FeffermanPNAS2014,ThorpPRA2016,LevyEPJST2017}, started to attract attention. In particular, it has been proven that 1D Schr\"odinger systems, in the presense of a domain wall modulated potential, possess topologically protected bound states, emerging from an effective 1D Dirac equation~\cite{FeffermanPNAS2014}. Similarly, the existense and robustness of topologically protected guided wave modes in a domain wall modulated photonic crystal are investigated in Ref.~\cite{ThorpPRA2016}. Interestingly, the existense of interface states has been related to the connection between the surface impendance and the Zak phases of the bulk band structure~\cite{XiaoPRX2014,ChoiOptLett2016}. This finding has been experimentally demonstrated for acoustic waveguides~\cite{XiaoNatPhys2015}.

In this work, we study the impact of finite size on the properties of one-dimensional scattering systems which support interface states 
emerging from a topological phase transition of the corresponding infinite periodic counterpart.  After reviewing the properties of 
infinite periodic photonic crystals which feature a topological transition, we show how the band inversion is manifested in the scattering properties of the finite system. We establish an one-to-one correspondence between the Dirac point of the infinite system and a perfect transmission resonance of the unit cell. A fingertip of  band inversion is also found in the distribution of the poles of the scattering matrix (complex resonances frequencies~\cite{Pangeux2013}), since the two relevant bands exchange a pole. 
To understand the genesis of this interface state we suitably vary the refractive  index of an initially periodic finite structure and follow the trajectory of the corresponding poles of  the scattering matrix.
To further exploit the role of finite effects we study how the interface state is affected while  the size of the system
is reduced and we surprisingly observe that the associated mode is still evident in the smallest structure possible i.e. one unit-cell
in either side of the interface.

The paper is organized as follows: In Sec.~\ref{sec2} we remind the basic notions of topological interface states in infinite periodic 
systems. In Sec.~\ref{subsec3_1}  we study the manifestation of band inversion in
the transmittance and the complex resonance (pole of the scattering matrix) distribution of a finite periodic crystal.
In Subsec.~\ref{subsec3_2} we follow the generation of the interface state by suitably varying the refractive indices of the structure
 and finally in Subsec.~\ref{subsec3_3} we study the fate of the interface state as the system size decreases.

\section{Reminder on Topological interface states in infinite systems }\label{sec2}
\begin{figure}[h!]
\includegraphics[width=0.9\columnwidth]{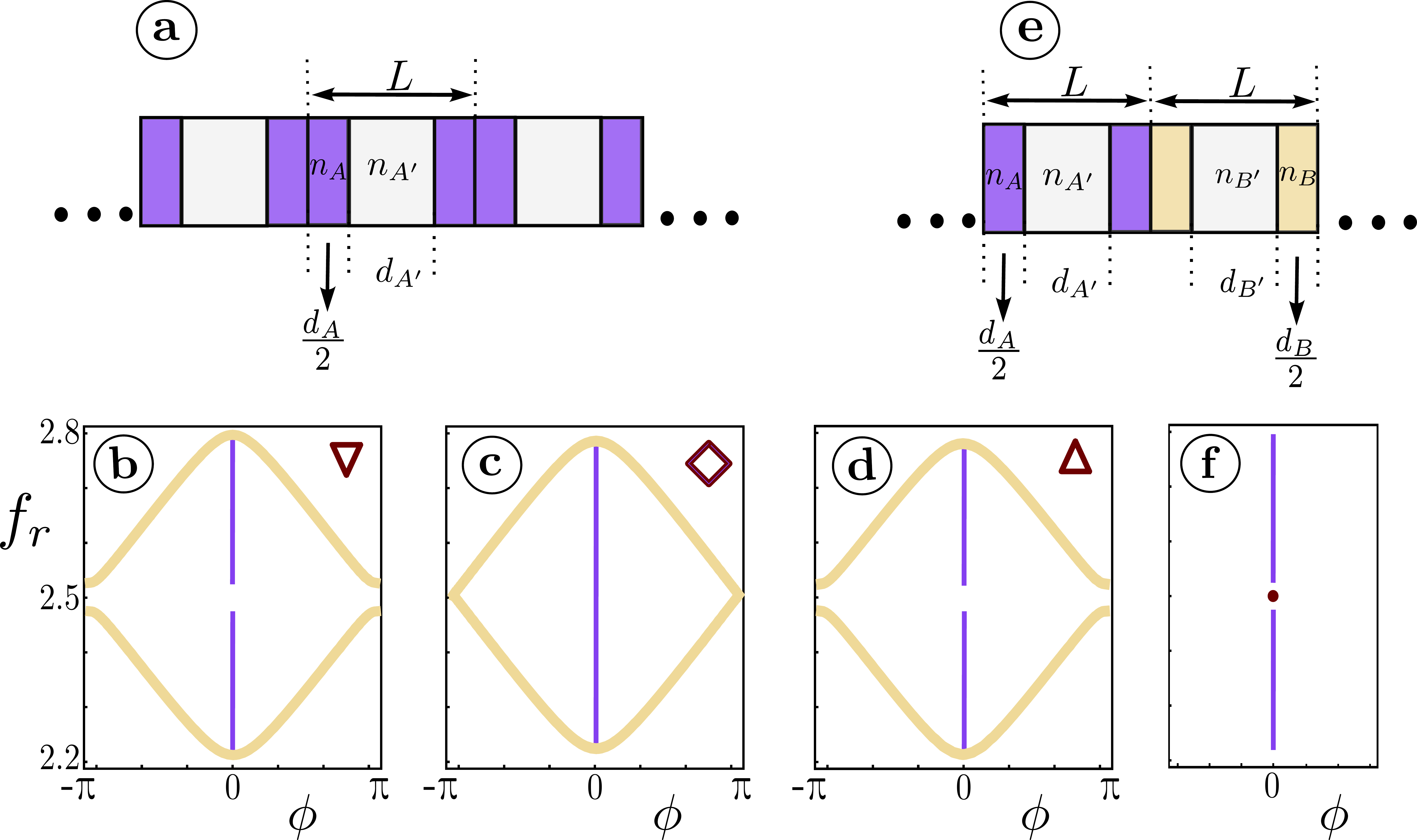}
\caption{(Color online) (a) Schematic of an one-dimensional infinite photonic crystal. (b), (c), (d) Band structure of the photonic crystal 
for  $n_{A}=1.95$, $2$, and $2.05$ respectively. 
(e) Schematic of two attached semi-infinite periodic systems, the left one with $n_A=1.95$ and the right one with $n_B=2.05$. (f) The spectrum of (e) with an interface mode inside the gap. }
\label{fig2} 
\end{figure}

In this Section we remind the properties of topological interface states for the case of 
an infinite photonic crystal.
In particular, we consider a 1D photonic crystal, as shown in Fig.~\ref{fig2} (a), which is composed by alternating slabs  with
refractive indices $n_{A},~n_{A'}$ and lengths $d_{A},~d_{A'}$, respectively. The unit cell has a length $L=d_{A}+d_{A'}$ 
and the corresponding dispersion relation is given by~\cite{Yariv1984}

\begin{align}
\label{per_band} \cos(\phi) &= \cos(k_{A} d_{A}) \cos(k_{A'} d_{A'}) \nonumber \\ &-   \frac{1}{2} \left( \frac{k_{A}}{k_{A'}}+ \frac{k_{A'}}{k_{A}} \right) \sin(k_{A} d_{A}) \sin(k_{A'} d_{A'}), 
\end{align}
where $\phi=\kappa L$ is the Bloch phase, $\kappa$ the Bloch wave number and $k_{i}=2\pi n_{i} f,~(i=A,~A')$. In the above expression and for the rest
of this work we consider normalized units where length is normalized with the unit cell size i.e. $L=1$ while $c=1$ 
is the speed of light in vacuum.
Such a structure was recently studied in Ref.~\cite{XiaoPRX2014} where the correspondence between the surface impedance
and the Zak phase was established,  by linking the bulk bands (through the Zak phase) with the surface. This ``bulk-interface'' correspondence has proven to be a valuable tool for the determination of possible interface states in 1D multilayered structures. 
Here we review the three key findings of~\cite{XiaoPRX2014} which are essential for the rest of our analysis: 
\begin{enumerate}[i]
\item \textit{Band crossing, Dirac points and band inversion:} Considering the photonic crystal shown in Fig.~\ref{fig2} (a) it is possible to close and re-open a gap in the band structure by a suitable tuning of the refractive index (i.e. $n_{A}$). When the gap closes the bands cross linearly, forming a Dirac point. The re-opening of the gap can be accompanied by a change in the sign of the surface impedance and a switch in the Zak phase value between the two crossing bands. This corresponds to band inversion and signifies a topological phase transition.

\item \textit{Conditions for the band crossing:} 
It has been shown that there is a necessary and sufficient condition for the band crossing. This is valid if the ratio $\alpha=n_{A}d_{A}/n_{A'}d_{A'}$ of the optical paths of two slabs $A,~A'$ of a photonic crystal is a rational number, i.e., $\alpha=m_{A}/m_{A'}$ where $m_{A},~m_{A'}$ are positive integers. Then, the $(m_{A}+m_{A'})$-th and the $(m_{A}+m_{A'}-1)$-th bands will cross at the frequency $f=(m_{A}+m_{A'})c/2\tau$, where $\tau=n_{A}d_{A}+n_{A'}d_{A'}$ is the phase delay in each unit cell. Moreover, at the Dirac point  $\sin(k_{A}d_{A})=\sin(k_{A'}d_{A'})=0$ and $\cos(k_{A}d_{A})=(-1)^{\ell m_{A}},~\cos(k_{A'}d_{A'})=(-1)^{\ell m_{A'}},~\ell \in \mathbb{N}^{+}$. Thus, at the Dirac point the Bloch phase satisfies the relation $\cos(\phi)=\pm 1$ depending whether the crossing occurs at the center ($+$) or  the edge ($-$) of the Brillouin zone.

\item \textit{Existence of interface states:} When two semi-infinite photonic crystals with surface impedance of opposite signs 
and common gap frequency are brought into contact, the structure exhibits a localized interface state with frequency inside the gap. 
One possible way is when one of the photonic crystal corresponds to the band inverted case of the other,  which as we mentioned above it reached by tuning the system parameter in order to cross a topological phase transition.

\end{enumerate}

In the following we will use the same set of parameters as in Ref.~\cite{XiaoPRX2014}, which satisfies the aforementioned
properties. In particular, by choosing $n_{A}=2,~n_{A'}=1,~d_{A}=0.4,~d_{A'}=0.6$ a Dirac point is ensured  between the 
6-th and 7-th band of the photonic crystal at the normalized frequency  $f= 2.5$. Note that for these parameter values
$\tau=1.4$.
In  Fig.~\ref{fig2} we plot the part of the corresponding  band structure of the system
(in particular the 6-th and 7-th band) for three different values of the refractive index $n_{A}$.
Panels  (b), (c) and (d) correspond to the values $n_{A}=1.95,~2,~2.05$  respectively ($n_{A'}=1$ in all cases).
Starting from  $n_{A}=1.95$ and increasing the refractive index, the gap which separates the 6-th and 7-th bands decreases and the two bands cross at the Dirac point for $n_{A}=2$ as expected by the aforementioned argumentation. As $n_{A}$ is further increased the gap re-opens  and band inversion occurs.  
The formation of the Dirac point at $n_{A}=2$ and the band inversion signify a topological phase transition which is characterized by the Zak phase~\cite{ZakPRL1989} and its relation to the surface impedance of the photonic crystal in the corresponding gap. Note that in order to keep all midgap positions intact, the quantity $\tau$ has to remain constant as $n_{A}$ varies, meaning that $d_{A}$ has to change according to the relation
\begin{equation}
\label{dist} d_{A}=(\tau-n_{A'})/(n_{A}-n_{A'}).
\end{equation}

The appearance of an interface mode is then ensured by considering  two semi-infinite periodic systems. At left, the first system has a unit cell $\mathcal{A}$ with refractive index values $n_{A}=1.95,~n_{A'}=1$, such that the gap between the 6-th and 7-th lies before the band inversion has taken place [Fig.~\ref{fig2} (b)].
At right, the second semi-infinite system has a unit cell $\mathcal{B}$ of refractive index values $n_{B}=2.05,~n_{B'}=1$ and the 
gap between its 6-th and 7-th bands lies in the region after band crossing [see Fig.~\ref{fig2} (d)].
The concatenation of these two semi-infinite periodic systems, leads to an infinite structure with an interface at the
connection point of the slabs $A$ and $B$,  as shown in Fig.~\ref{fig2} (e).
For this case, it has been shown that an interface mode exists which originates from the Dirac 
point at $f=2.5$~\cite{XiaoPRX2014,FeffermanPNAS2014} [see Fig.~\ref{fig2} (f)].

\section{Fate of topological interface states in 1D open finite systems}\label{sec3}
\begin{figure}
\includegraphics[width=0.9\columnwidth]{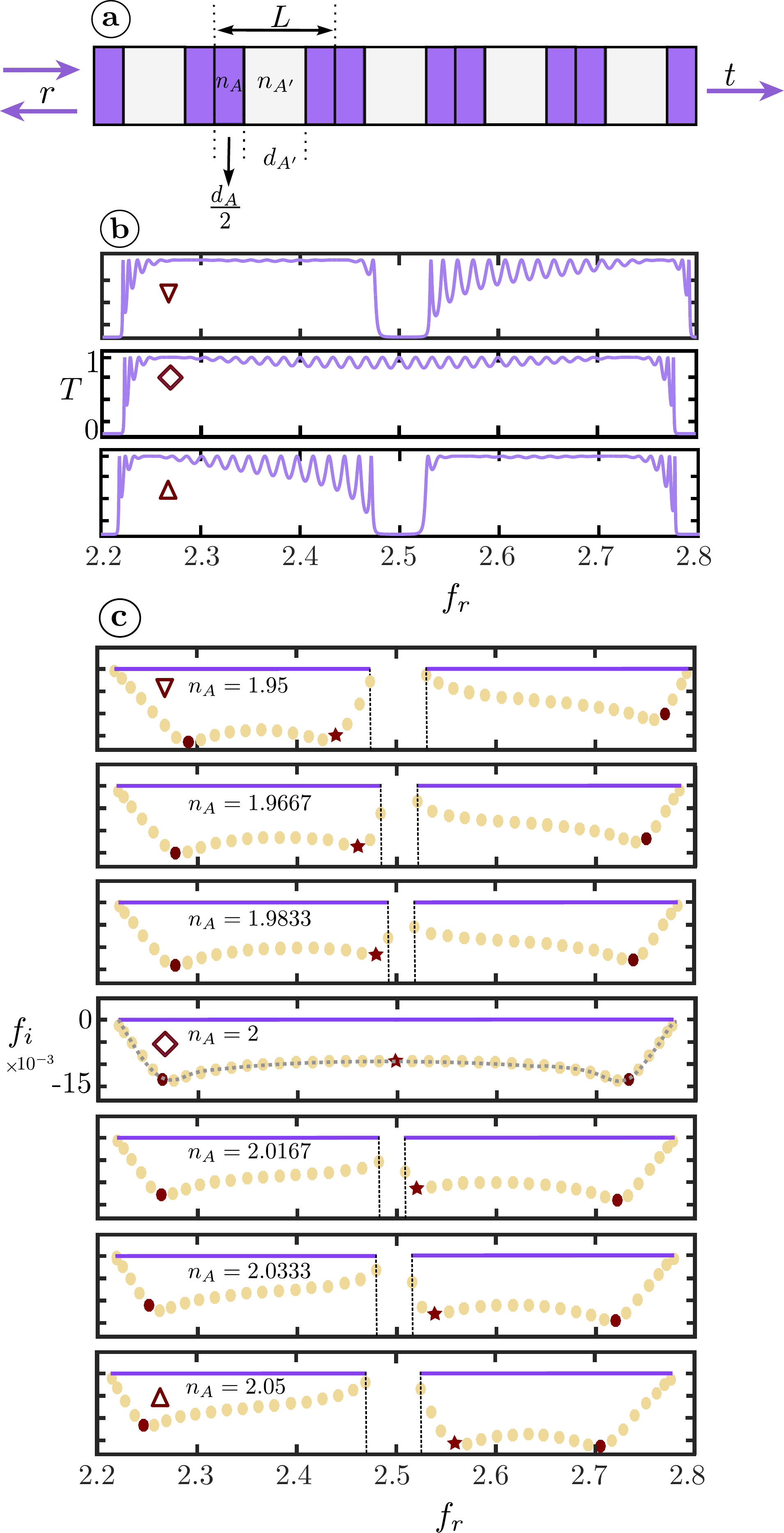}
\caption{(Color online) (a) Schematic of the scattering through a 1D finite periodic photonic crystal.
(b) Panel of plots illustrating the transmittance $T$  from a periodic structure of 20 cells with refractive index values
$n_{A}=1.95$ (point down triangle), $n_{A}=2$ (square), $n_{A}=2.05$ (point up triangle).
(c) Panel of plots showing the distribution of the poles of the scattering matrix when the refractive index $n_{A}$ varies
within the range $[1.95,~2.05]$ (top to bottom).
The three colored poles (circles and star) correspond to PTR of the corresponding unit cell.
The star pole is the one which is related to the origin of the interface state. The purple solid lines at $f_{i}=0$ correspond to the continuum spectrum of the infinite system.}
\label{fig3}  
\end{figure}

After the reminder on the properties of an interface state formed by connecting two semi-infinite
photonic crystals we now study the formation of complex resonances for a finite system as shown in Fig.~\ref{fig3} (a). 
To do so, we characterize the structure by its corresponding scattering and transfer matrices.

\subsection{Band inversion on the complex plane}\label{subsec3_1}

It is instructive to first study the transfer matrix of a unit cell which can be used to derive the dispersion
relation of the infinite periodic structure but also holds information about the finite system. The unit cell
is parity symmetric with an inversion center at the middle of the slab with refractive index $n_{A'}=1$ as it is indicated in
Fig.~\ref{fig3} (a).
The  transfer matrix of the unit cell is thus given by
\begin{equation}
M=
  \begin{bmatrix}
   1/t_{1} & r_{1}^{*}/t_{1}^{*} \\
    r_{1}/t_{1} & 1/t_{1}^{*}
  \end{bmatrix},
\label{tm} 
\end{equation}
where $r_{1},~t_{1}$ are the reflection and transmission coefficients  respectively and $^*$ denotes complex conjugation.
For the photonic crystal considered here, the transfer matrix is obtained using

\begin{align}
M=\displaystyle \prod_{i=1}^{4} M_i(k_i,k_{i+1},x_{i})
\label{matrixunit}
\end{align}

with 
\begin{align}
M_i=\small \frac{1}{2k_{i}} \begin{bmatrix}
(k_{i}+k_{i+1}) e^{i(k_{i+1}-k_{i})x_{i}} & (k_{i}-k_{i+1}) e^{-i(k_{i+1}+k_{i})x_{i}} \\
(k_{i}-k_{i+1}) e^{i(k_{i+1}+k_{i})x_{i}} & (k_{i}+k_{i+1}) e^{-i(k_{i+1}-k_{i})x_{i}}
\end{bmatrix}.
\label{matrixunit1}
\end{align}

with $k_{1,3,5}=2\pi n_{A'} f$, $k_{2,4}=2\pi n_{A} f$ and $x_1=0$, $x_2=d_A/2$, $x_3=d_A/2+d_{A'}$ and $x_4=d_A+d_{A'}$.

The Bloch phase can be obtained from the transmission of the  unit cell via the relation,
\begin{equation}
\label{bloch_t} \cos\phi=\textrm{Re}\left[\frac{1}{t_{1}}\right],
\end{equation}
which recovers Eq.~(\ref{per_band}). This equation  provides a direct link between the infinite system (Bloch phase)
and the finite structure (transmission coefficient $t_1$).
Additionally, using the transfer matrix of the unit cell, Eq.~(\ref{tm}), we obtain the transmittance $T=|t_N|^2$  for
a system of N cells given by~\cite{SprungAJP1993,AchilleosPRL2017}
\begin{equation}
\label{finite_per_T} T^{-1}=1+\frac{\sin^{2}N\phi}{\sin^{2}\phi}\left(\frac{1}{|t_1|^2}-1\right).
\end{equation}

The transmittance $T$, from a finite periodic crystal of 20 cells, is shown in Fig.~\ref{fig3} (b)  as a function 
of the the frequency $f_{r}$ for refractive index values $n_{A}=1.95,~2,~2.05$ corresponding to the panels (b), (c) and (d) of  Fig.~\ref{fig2}. The red symbols emphasize this correspondence. 
Note that the transmittance is characterized by a sequence of perfect transmission resonances (PTRs) where $T=1$ within the 
propagating bands.  From Eq.~(\ref{finite_per_T}) it appears that within each band there
exist  $N-1$ peaks~\cite{SprungAJP1993}, stemming from Bragg reflections at frequencies 
satisfying  $N\phi=m\pi$ with $m=1,2\ldots,N-1$.
Besides, additional PTRs of the finite 
periodic system appear at frequencies where the unit cell transmittance is $|t_1|^2=1$.

For example, in the case of $n_A=1.95$ and for a system of $N=20$ cells, there exist 19 PTRs due to Bragg reflections in each pass band. 
For the same parameters,  the transmission of the unit cell in the range of frequencies $f_{r} \in [2.2,2.8]$ exhibits two PTRs in the left 
and one PTR  in the right band. Thus the total number of PTR peaks in the left and right bands of Fig.~\ref{fig3} (a) are 21 and 20 
respectively. During the band inversion, we observe that one complex resonance corresponding to a PTR of the unit cell (denoted by a star) is exchanged between the two bands.  It appears that after the band inversion, for $n_A=2.05$, the number of PTRs at each of the two bands is reversed i.e. we
count 20 and 21 peaks in the left and right bands respectively. 
The above described exchange of PTRs during band inversion is directly visualized  when we calculate  
the poles of the scattering matrix in the complex plane, which will be proven to be crucial to identify the origin of the interface states.
It has been shown in Ref. ~\cite{BarraJPhysA1999} that for a finite periodic crystal there is an one-to-one correspondence between the number of PTRs and the number of the scattering matrix poles.

The plots in Fig.~\ref{fig3} (c) illustrate pole distribution as the the refractive index varies
from $n_{A}=1.95$ to $n_{A}=2.05$, in the complex frequency plane where $f_i$ and $f_r$ denote the imaginary and real parts of the frequency respectively. The yellow circles correspond to the PTRs which 
emerge from the Bragg reflections i.e. the sinusoidal term of Eq.~(\ref{finite_per_T}). The red circles and the star correspond to the 
PTRs of the unit cell.  For the first, middle and last sub-panels, the refractive indices are the same with those at Figs.~\ref{fig2} (b), (c), 
and (d) respectively. The purple solid lines at $f_{i}=0$ indicate the corresponding bands of the continuous spectrum obtained from Eq.~(\ref{per_band}) for a 
more direct comparison between the finite (poles) and infinite systems (bands). Note that in the periodic case, all 
poles (both Bragg and unit-cell resonances) are distributed  within the bands of the infinite system, and the numbering of
the poles per band follows the aforementioned numbering of the PTRs.
\begin{figure}
\includegraphics[width=0.9\columnwidth]{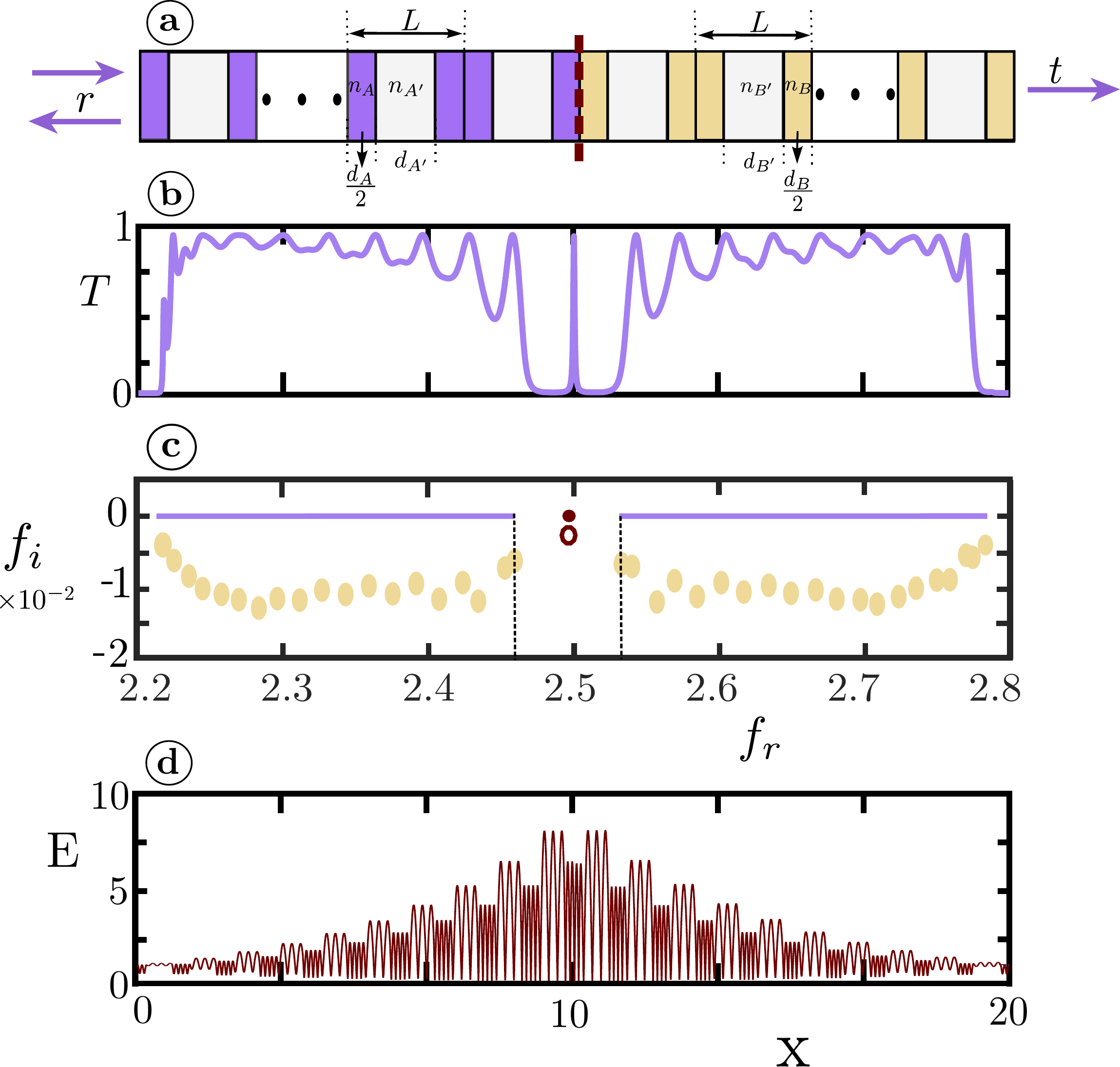}
\caption{(Color online) (a) The finite counterpart of Fig.~\ref{fig2} (e). (b) The transmittance of the system.
 (c) Distribution of the scattering matrix poles in the complex frequency plane. With the solid purple line and the red dot we present the continuous spectrum of the infinite system. The hollow red circle is the pole corresponding to the interface state.
 (d) The magnitude wave field of the interface state. $x$ is measured in unit cell lengths $L=1$.}
\label{fig4} 
\end{figure} 

As $n_{A}$ increases [top to bottom in Fig.~\ref{fig3} (c)], the distributed poles in the two bands approach each other
following the decrease of the corresponding gap width. The distribution in each band is different, with the left band featuring
poles with larger imaginary parts (more leaky) and two characteristic minima. In fact, the minima of both bands correspond to the 
states which are induced by the PTRs of the unit cell, as the red colored poles indicate. While $n_{A}$ increases we focus on the motion of the pole denoted with a star. As the refractive index varies and the gap closes this pole moves towards the band edge. At $n_{A}=2$ where the band crossing of the infinite system occurs 
[see Fig.~\ref{fig2} (c)], this pole is located at the Dirac frequency $f_{r}=2.5$ and the distribution becomes
symmetric in the domain spanned by the two merged bands. The corresponding PTR at the Dirac point is also directly
seen from Eq.~(\ref{bloch_t}), since at this point $\phi=\pi$, and thus $t_1=-1$. After the band inversion (at $n_{A}=2$) this pole 
passes to the band on the right. 
Thus we have shown that, in a finite periodic system the band inversion is manifested by an exchange of a pole of the scattering
matrix, between the inverted bands, and that this pole corresponds to a PTR of the unit cell.
Following this exchange the shape of the distribution of the poles in the two bands is also exchanged,
as readily seen by  comparing the first and last plots of Fig.\ref{fig3} (c). 
We note that by choosing any other non parity symmetric unit cell the resulting finite photonic crystal will not be parity symmetric
and this leads to a transmittance and pole distribution different from the one shown in  Fig.~\ref{fig3}. However, band inversion
is always associated with the exchange of one PTR of the unit cell passing through the Dirac point.

More information regarding the finite  periodic case at the Dirac point, can be obtained using  
Eqs.~(\ref{per_band}),~(\ref{tm}) and (\ref{matrixunit}), by noting that at $\phi=\pi$ we 	
find $r_{1}=0$ and $t_{1}=-1$. Thus, scattering through the unit cell at the Dirac frequency results just in a phase acquisition equal to $\pi$. 
Actually, the fact that the transmission amplitude of the unit cell is $t_{1}=-1$ has interesting implications on the transmission amplitude $t_{N}$ of the $N$-cell finite periodic system, which is given by the equation,  

\begin{equation}
\label{finite_per_t_ampl}  t_{N}^{-1}=\frac{1}{t_{1}} \frac{\sin N\phi}{\sin\phi}-\frac{\sin(N-1)\phi}{\sin\phi}.
\end{equation}

Thus, for an even number $N$ of cells (at $\phi=\pi$), $t_{N}=1$ rendering the system invisible, meaning that after a scattering experiment (at the Dirac point parameters), the outgoing wave  will carry no information about the scattering process. On the other hand, for an odd number $N$ of cells  $t_{N}=-1$, which implies a phase acquisition equal to $\pi$.
Note that if the band crossing in the corresponding infinite periodic system occurred at the center of the Brillouin zone, i.e. at $\phi=0$, then it would hold that $t_{1}=1$, rendering the system invisible. 

An interface state can be formed by bringing
together two finite periodic structures, e.g.  one with $n_A=1.95$ (before  band crossing) and one with $n_A=2.05$ after band crossing,
as shown schematically in Fig.~\ref{fig4} (a). The effect of the interface is directly seen in the tranmsittance for a system 
composed by totally 20 cells, Fig.~\ref{fig4} (b), as a transmission peak appears within the common gap of the two
finite, 10-cell, periodic sub-systems. Additionally the pole distribution, shown in Fig.~\ref{fig4} (c), is significantly
altered compared to the periodic case, but it is still enclosed within the bands of the corresponding inifinite system. 
However there is a single pole at the Dirac frequency which corresponds to the interface mode (hollow circle) 
and is the remnant of the topological state of the infinite system (red dot). In Fig.~\ref{fig4} (d) we  show the
distribution of the field within the system where we observe that it is localized around the interface between the two 
photonic crystals.

Note that we have confirmed the presence of an interface mode for different non parity symmetric 
photonic crystals which are concatenated and these modes exhibit similar  frequencies, transmission amplitudes and field 
distributions.

\begin{figure}
\includegraphics[width=0.9\columnwidth]{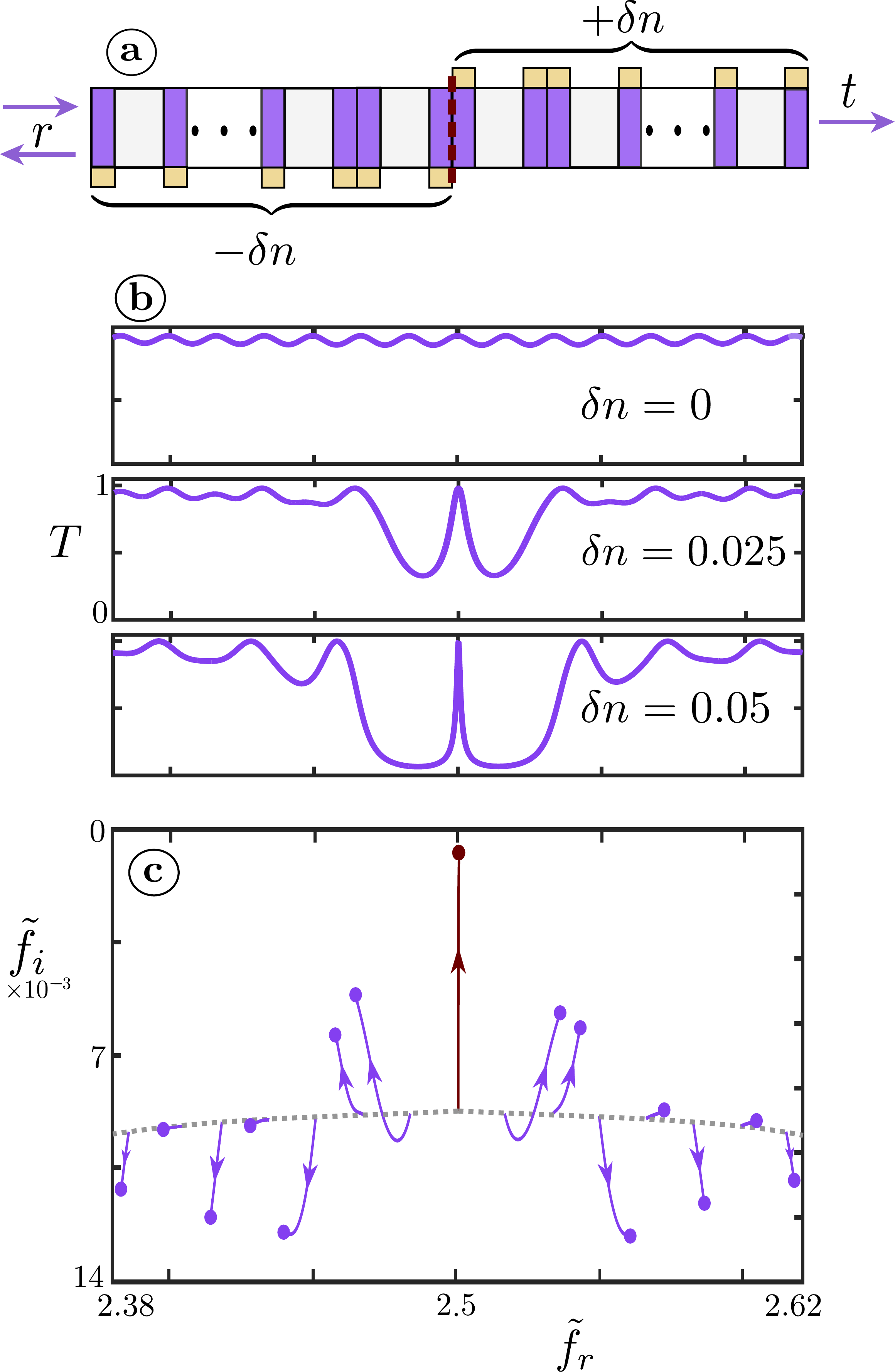}
\caption{(Color online) (a) Schematic of the how a simultaneous decrease (increase) from the left (right) of the refractive index $n_{A}$, from
an initially periodic structure forms an interface.
(b) Panel of transmittance plots for three different values of $\delta n$ ($\delta n =0$,  $\delta n =0.025,~\delta n=0.05$).
(c) Trajectories of the poles of the scattering matrix in the complex frequency plane as the $\delta n$ increases up to the value 0.05. The 
Dotted line depicts the distribution at the periodic case ($\delta n=0$). The pole corresponding to the interface mode (central) is shown with red color.}
\label{fig5} 
\end{figure} 
 
\subsection{Towards the interface state}\label{subsec3_2}
 
To study the formation of the interface state, we start from a periodic structure of 20 cells with $n_{A}=2$,
and simultaneously decrease the refractive index  $n_{A}$ of 10 unit cells on the left-side and increase 
$n_A$ of the 10 unit cells in the right-side by the same amount $\delta  n\in [0,0.05]$. In such a way, as shown
in Figure~\ref{fig5} (a), we create a structure 
composed by two concatenated photonic crystals with overlapping band gaps before and after the band crossing (as in Fig.~\ref{fig2}).
As $\delta n$ is increased we observe that the transmittance of the whole system forms a gap around
Dirac frequency $f=2.5$ and a  transmission peak remains in the center of this gap. The formation of this gap -as we start from a finite periodic system  and $\delta n$ increases- is illustrated in panel Fig.~\ref{fig5} (b).
We next explore the connection of this particular PTR (and the corresponding pole) with the formation of an interface state when the system departs from periodicity. 
The direct link between the PTR and the interface state, is established in Fig.~\ref{fig5}(c) where we plot the trajectories of the  poles of the 20 cell system, as  $\delta n$ increases. The arrows indicate the trajectory starting from the periodic case
 with $\delta n=0$ (dashed line) to a final value of $\delta n=0.05$ (purple points).

It is then seen that as we deviate from the periodic case, all poles drift away from the Dirac point frequency $f=2.5$, except for the red colored pole which remains at this point.
In fact the drifting of the poles forms a band gap as shown in Fig.~\ref{fig5} (c) and the central pole associated with the PTR of the unit cell forms an interface state in the middle of this gap. Additionally, following the trajectory of the pole corresponding to the interface state, we observe that it approaches the real axis, as both $\delta n$ and the band gap increase, and thus becomes narrower (less leaky). This is obvious also from the transmission spectra in panel (b) of Fig.~\ref{fig5}. Using the above description with the poles of the scattering matrix, we thus identified the origin of the interface state as a PTR of the unit cell at the Dirac point.

\begin{figure*}
\includegraphics[width=0.9\textwidth]{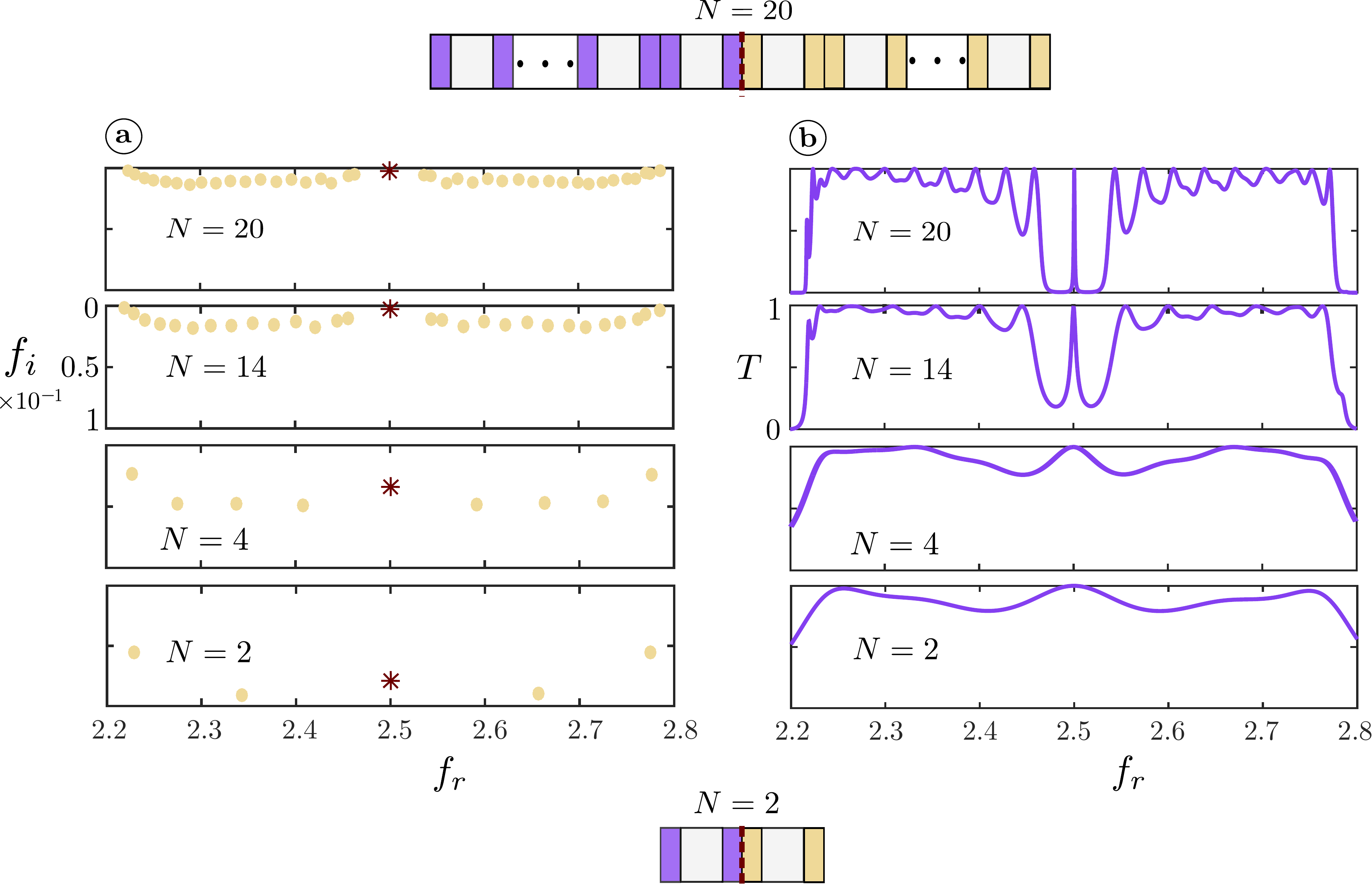}
\caption{(Color online) Size variation of a structure with an interface state. The size varies from a total number of 20 cells to the minimum possible number of two cells. (a) Scattering matrix pole distribution for four different system sizes. As the size decreases the pole bands depopulate (yellow dots). However, the interface mode, shown with the red star, remains almost unaffected with respect to $f_{r}$. The imaginary part of the frequency $f_{i}$ decreases, rendering the state more leaky. (b) Transmittance counterpart of (a). For large $N$, the interface state in the center of the gap is more pronounced and the gap more well defined. For smaller sizes the peak becomes wider implying a smaller quality factor. Nevertheless, it is unaffected with respect to the $f_{r}$, even for the smallest possible case of two cells.}
\label{fig6} 
\end{figure*}

\subsection{Size variation effect}\label{subsec3_3}
Having established the connection between a finite periodic system and the formation of an interface state,
we naturally proceed by studying  how the interface mode is affected when the size of the system varies. 
To do so, we start from the finite structure composed of 20 cells (with parameters corresponding to the system discussed in  Fig.~\ref{fig5} (b) for   $\delta n=0.05$). Then, by removing one cell from each end of the structure at the same time we gradually decrease the size of the system down to a two cell structure which still possesses an interface.  This reduction is schematically
shown in  Fig.~\ref{fig6}.
In particular, Fig.~\ref{fig6}  (a) shows the scattering matrix pole distribution for four different descending system sizes. As expected, during the reduction of the system size the pole distribution is significantly altered and the bands depopulate.  However, we observe an unexpected robustness of the pole which corresponds to the topological interface mode  (shown with the red star) since it remains unaffected with respect to the $f_{r}$ axis, being always in the vicinity around $f_r=2.5$.  This  signals the appearance of an interface state residing in the vicinity of the band gap of the periodic system, even if there is only one unit cell from each side of the interface.

This characteristic is also reflected on the transmittance of the system (Fig.~\ref{fig6} (b)). The number of peaks and their widths vary as the system size decreases. There is a peak, however, which always resides inside the band gap region around the same $f=2.5$ value and this corresponds to the interface state and the red star in the pole distribution. For the larger system, the peak of the interface state becomes more pronounced since the gap begins to form, while, the smaller systems possess a wider interface state peaks and the gap is not clearly distinguished. This can be understood from the interface mode in the complex plane which, while it remains at the same $f_{r}$, its imaginary frequency $f_{i}$ decreases, implying a more leaky mode and a small quality factor.  Nevertheless, the existence and the position of the peak is not affected by the size of the finite composite structure.

This striking robustness is not necessarily expected. The emergence of topological interface modes in scattering setups is a signature of the topological phase transition which occurs in infinite periodic systems. However, here we find that this mode is not only persistent in large finite structures which carry the properties of their infinite counterparts, but it manifests in the same $f_{r}$ even for the smallest possible realization of the system (with only 2 cells). 

\section{Conclusions}\label{sec5}

We have studied the effects of finite size on one-dimensional photonic crystals exhibiting interface states which emerge from a 
topological phase transition of the corresponding infinite periodic system. We found that for the finite periodic system the Dirac point 
manifests as a perfect transmission resonance which is induced by the unit cell. We also concluded that the band inversion in a finite 
periodic system is indicated by the exchange of a unit cell perfect transmission resonance between the two bands. 
Then we showed that as the system departs from periodicity and the interface is formed, the mode which corresponds to the interface 
state is emerging directly from a perfect transmission resonance of the unit cell. Finally, we reported on the striking robustness of the 
interface state as the system size varies by showing  that it persists almost at the same frequency for the smallest possible system, even 
if its generating property - the band crossing- occurs in the infinite system. Apart from the theoretical  insight regarding the origin of 
these states, the finite size of the systems discussed renders our findings relevant to realistic and possibly experimentally realizable 
scattering structures.

\begin{acknowledgments}
This work was financially supported by Le Mans Acoustique in the framework of the APAMAS project. P.A.K would like to thank F. Diakonos for illuminating discussions.
\end{acknowledgments}

\end{document}